\documentclass[11pt]{article}
	
	\newcommand{\blind}{0}
	
    \makeatletter
    \renewcommand\section{\@startsection {section}{1}{\z@}%
                                       {-3.5ex \@plus -1ex \@minus -.2ex}%
                                       {2.3ex \@plus.2ex}%
                                       {\normalfont\fontfamily{phv}\fontsize{14}{17}\bfseries}}
    \renewcommand\subsection{\@startsection{subsection}{2}{\z@}%
                                         {-3.25ex\@plus -1ex \@minus -.2ex}%
                                         {1.5ex \@plus .2ex}%
                                         {\normalfont\fontfamily{phv}\fontsize{12}{15}\bfseries}}
    \renewcommand\subsubsection{\@startsection{subsubsection}{3}{\z@}%
                                        {-3.25ex\@plus -1ex \@minus -.2ex}%
                                         {1.5ex \@plus .2ex}%
                                         {\normalfont\normalsize\fontfamily{phv}\fontsize{12}{15}\selectfont}}
    \makeatother
	\usepackage{amsmath}
	\usepackage{graphicx}
	\usepackage{enumerate}
	\usepackage{xcolor}
	\usepackage{float}
	\usepackage[square,comma,numbers,sort&compress]{natbib} %comment out if you do not have the package
	\usepackage{url} % not crucial - just used below for the URL
\begin{document}
		
			%%%%%%%%%%%%%%%%%%%%%%%%%%%%%%%%%%%%%%%%%%%%%%%%%%%%%%%%%%%%%%%%%%%%%%%%%%%%%%
		\def\spacingset#1{\renewcommand{\baselinestretch}%
			{#1}\small\normalsize} \spacingset{1}
		%%%%%%%%%%%%%%%%%%%%%%%%%%%%%%%%%%%%%%%%%%%%%%%%%%%%%%%%%%%%%%%%%%%%%%%%%%%%%%
		
		\if0\blind
		{
			\title{Synchronization based model for turbulent thermoacoustic systems}
			\author{Yue Weng$^1$, Vishnu R. Unni$^{1,2,\#}$, R. I. Sujith$^{3}$, Abhishek Saha$^{1,*}$ 
			\\
			\\
			\small
			$^1$ Department of Mechanical and Aerospace Engineering, University of California San Diego, \\
			\small La Jolla, CA-92093, USA 
			\\
            \small
            $^2$ Department of Mechanical and Aerospace Engineering, Indian Institute of Technology Hyderabad, \\
            \small Kandi, Sangareddy-502284, India
            \\
            \small
            $^3$ Department of Aerospace Engineering, Indian Institute of Technology Madras, \\
            \small Chennai-600036, India \\
            \small
            emails: $^\#$vishnu.runni@mae.iith.ac.in $\&$ $^*$asaha@eng.ucsd.edu}
			\date{}
			\maketitle
		} \fi
		
		\if1\blind
		{

            \title{\bf \emph{IISE Transactions} \LaTeX \ Template}
			\author{Author information is purposely removed for double-blind review}

\bigskip
			\bigskip
			\bigskip
			\begin{center}
				{\LARGE\bf \emph{IISE Transactions} \LaTeX \ Template}
			\end{center}
			\medskip
		} \fi
		\bigskip
		
	\begin{abstract}
We present a phenomenological reduced-order model to capture the transition to thermoacoustic instability in turbulent combustors. 
The model is based on the framework of synchronization and considers
the acoustic field and the unsteady heat release rate from turbulent reactive flow as two nonlinearly coupled sub-systems.
Previous experimental studies have reported a route from low amplitude chaotic oscillation (i. e. combustion noise) to periodic oscillations through intermittency in turbulent combustors.
By varying the coupling strength, our proposed model can replicate the route that is observed in experiments.
Instead of assessing combustion noise as background noise to the system, the model considered a coupled oscillator system that produces multifractal chaotic oscillations to represent the combustion noise. 
This set of coupled oscillators is then nonlinearly coupled to a linear oscillator representing the acoustic field. 
As the coupling strength increases, the system loses its multifractality and exhibits large-amplitude periodic oscillations in a manner consistent with the dynamics observed in experimental studies.
	\end{abstract}
			
% 	\noindent%
% 	{\it Keywords:} \emph{IISE Transactions}; \LaTeX; Manuscript format; Taylor \& Francis.

	%\newpage
	\spacingset{1.5} % DON'T change the spacing!

\section{\label{Intro}Introduction}
In most power generation devices, including gas turbines, rocket engines, and industrial burners, the combustion process usually occurs in an enclosed space. %with high energy densities.
In these devices, the heat release rate fluctuations may interact with the acoustic field in the combustor, and vice versa.
As a result, feedback may arise, and when this feedback is positive, there will be growth in acoustic fluctuations,
which eventually saturates to limit cycle oscillations due to nonlinearities in the system. This phenomenon is known as thermoacoustic instability.
The violent oscillations are usually harmful to combustion devices, they can cause the extinction of the flame, or damage the hardware~\cite{Lieuwen_book_2005, Oefelein_JPP_1993, poinsot_JFM_1987, candel_Symposium_1992}. 
In the past few decades, lean-premixed-prevaporized (LPP) combustors are widely applied in the industry for reducing emissions. %In these combustors, however, 
The flame is inherently more sensitive to perturbation when operating in fuel-lean conditions. Thus, the LPP combustors are more susceptible to thermoacoustic instability.
Because of the wide application of LPP combustors, thermoacoustic instability has become a major concern in combustor design. Numerous studies in the past have focused on prediction of onset, evasion or suppression of thermoacoustic instabilities.

Rayleigh first explained the emergence of thermoacoustic instability in 1878. % how acoustic energy is added due to heat release fluctuations
He showed that when the acoustic oscillation is in phase with the fluctuation of unsteady heat release rate, energy is added to the acoustic field, hence the amplitude of acoustic pressure will grow~\cite{Rayleigh_1878_Nature}.
The amplitude would keep increasing until limited by the nonlinear effects. 
To identify the regime where thermoacoustic instability occurs, investigating the feedback mechanism is of crucial importance~\cite{dowling1997jfm}. 
In many cases, the propagation of acoustic waves is considered to be a linear phenomenon since the amplitude of pressure oscillations corresponding to the perturbations is relatively low when compared to the mean pressure in the combustor~\cite{sujith2021book}. %\hl{[cite]}. 
The flame, however, is more complex due to the inherent nonlinearities associated with the chemical reactions and transport processes of heat and momentum~\cite{wu2003jfm}. The response of the flame to the acoustic perturbation is also highly nonlinear, especially in a turbulent environment~\cite{sujith2021book}. 

Traditionally, modeling of thermoacoustic instability aims to determine the response of the flame to acoustic perturbation.
In the past few decades, several low-order models based on quasi-linear approach were introduced. Among them, the most commonly used models are flame transfer function (FTF)~\cite{dowling1997jfm, candel2002proceeding, dowling2003JPP, ducruix2000proceeding, lieuwen2003jpp, cheung2003turbexpo} and flame describing function (FDF)~\cite{dowling1997jfm, stow2004turbexpo, noiray2008jfm, han2015cnf}. 
FTF estimates the linear response of the flame to acoustic perturbations by either experiment or CFD.
External velocity perturbation ($u'$) is introduced to the combustor by using actuators (speakers or linear actuators). 
Then, %defined in the frequency domain, 
the response in form of the fluctuation in the heat release rate ($\dot q'$) is measured and evaluated as a function of the forcing frequencies ($\omega$). The flame transfer function, $F$, is, then, defined as shown in Eq.~(\ref{eq: FTF}), where $\bar{\dot q}$ and $\bar u$ are the mean heat release rate and the mean velocity at a given condition.

\begin{equation}
\label{eq: FTF}
  F(\omega)~=~\frac{\dot q'(\omega)/\bar {\dot q}}{u'(\omega)/\bar u}~=~G(\omega)e^{i\varphi(\omega)}
\end{equation}
%\hl{Once by applying FTF as a forcing term, frequencies that result in a positive growth rate can be identified, giving us the stability regime of the system.}
%\hl{talk about acoustic equation, e.g. forcing term, write very specifically}
%\hl{Mar 2, Wednesday}
Once the FTF is estimated, the global dynamical behavior of a thermoacoustic system can be evaluated by incorporating the acoustic driving due to heat release rate fluctuations (obtained from FTF) into the acoustic field equations.
%of the acoustic field to heat release perturbations at different frequencies can be evaluated by incorporating the forcing due to heat release rate fluctuations into the acoustic field equations using the FTF.
However, established on classic linear stability analysis, FTF cannot capture the nonlinear effect of the thermoacoustic system.
Thus the amplitude of limit cycle oscillation, or triggering phenomenon in a subcritical Hopf bifurcation, cannot be estimated using FTF~\cite{strogatz_2018_book}.
Subsequently, flame describing function (FDF) was introduced. Originated from theories of control systems~\cite{vander1968MGL} and introduced by Dowling et al.~\cite{dowling1997jfm, dowling2003JPP}, FDF extended the framework of FTF to the nonlinear regime. It defines the complex response of the flame as a function of both input frequency ($\omega$) and amplitude ($|u'|$), as shown in Eq.~(\ref{eq: FDF}).

\begin{equation}
    \label{eq: FDF}
    F(\omega,|u'|)~=~\frac{\dot q'(\omega)/\bar {\dot q}}{u'(\omega)/\bar u}~=~G(\omega,|u'|)e^{i\varphi(\omega,|u'|)}
\end{equation}
In the FDF approach, the flame is considered as a nonlinear module coupled to the linear acoustic system. Compared to FTF, FDF can predict %model some weak nonlinearity in a thermoacoustic system, such as 
the amplitude of limit cycle oscillation. However, FDF is based on quasi-linear assumption, i.e. for a particular forcing, the flame only responds at the same frequency. In other words, the response of flame at the higher harmonic frequencies resulting from the forcing is ignored~\cite{juniper2018AR}. Also, FDF may not be suitable at modeling complex behaviors that include non-harmonic responses, e. g. chaos and intermittency. %quasi-periodic oscillation. 
Moreover, in FDF, the flame is forced by acoustic perturbations at different frequencies and amplitude, which similar to FTF, does not include many aspects of inherent physics of the flame dynamics~\cite{weng_2020_nody}. 
For a long time, the conditions of a thermoacoustic system were considered either operational stable or unstable (limit cycle thermoacoustic oscillation). Correspondingly, FTF/FDF approaches mainly focus on finding out the unstable regime and the amplitude of periodic oscillation.
However, experimental observations have shown that thermoacoustic systems can exhibit complex dynamical behaviors other than limit cycle oscillations~\cite{keanini1989asm, sterling_1993_CST, jahnke_1994_JPP}. Many of these behaviors can be identified from laminar thermoacoustic systems.
Beyond limit cycle oscillation, Kabiraj et al. ~\cite{kabiraj_2012_chaos} observed a route to chaos through quasi-periodic oscillation in a laminar Rijke tube, which conforms Ruelle-Takens scenario. 
Various bifurcations in the same system were subsequently investigated~\cite{kabiraj2012jegtp}, and complex behaviors including strange non-chaos were captured~\cite{premraj_2020_EPL}. 
These finding were also supported by other groups. For example, in a separate study, Guan et al.~\cite{guan2020jfm} also identified an intermittency route to chaos in a laminar Rijke tube.
Collectively, these studies show that even simple thermoacoustic system can exhibit rich dynamics.
%When increasing flow rate in the combustor, intermittent bursts of periodic oscillation with high amplitude can appear randomly among low amplitude aperiodic oscillations. If further increase the flow rate, the system will finally enter the regime of limit cycle oscillation. During the transition from combustion noise to limit cycle oscillation, the system is self-organized.

Inspired by the observations, the underlying physics of these complex behaviors in thermoacoustic systems attracts wide research interest. 
Existing approaches based on one-way coupling and forced response cannot provide plausible explanations for these phenomena.
However, an alternative viewpoint based on synchronization is getting attention recently.
Pawar et al.~\cite{pawar_2017_jfm} showed that the onset of thermoacoustic instability is a result of the mutual synchronization between the unsteady heat release rate and acoustic fluctuations. 
Depending on the dynamic states of the system, the two signals can be either synchronized or de-synchronized.
Later studies by Mondal et al.~\cite{mondal2017chaos} found various synchronization states including phase locking, intermittent phase locking, and phase drifting.
Moon et al.~\cite{moon2020chaos} presented a thermoacoustic system that consists of two coupled turbulent combustors, and investigated their mutual synchronization. Premraj et al.~\cite{premraj2021nody} investigated the effect of system parameters on the 
synchronization behaviors of coupled oscillators.

These researches paved the way for building new models for thermoacoustic systems using the framework of synchronization. 
%Raaj et al.~\cite{raaj2019chaos} introduced a synchronization framework for modeling an aeroelastic system. 
Godvarthi et al.~\cite{godavarthi_2020_chaos} found that while varying the coupling strength, a system of locally coupled grid of chaotic Rossler oscillators coupled globally to a chaotic Van der Pol oscillator can become self-organized and transition to periodic oscillations.
Guan et al.~\cite{guan2021PRE} introduced a low order model consisting of two Van der Pol oscillators.
In our recent work, we presented a phenomenological model based on the synchronization framework for modeling laminar thermoacoustic systems~\cite{weng_2020_nody}.
We use a pair of damped linear oscillators to represent the unsteady heat release rate and acoustic fluctuation, the two oscillators are nonlinearly coupled.
By adjusting the coupling strength, the model can replicate the Hopf bifurcation in a laminar thermoacoustic system, and capture the route to chaos, and also states including quasi-periodic oscillation, chaos, and strange non-chaos.
In the present work, we attempt to extend the synchronization framework for thermoacoustic systems to the turbulent regime. 

In turbulent combustors, turbulent reactive flow always generates low amplitude disordered acoustic fluctuations in the chamber. During stable operation, the sound is known as combustion noise.
Traditionally, combustion noise is considered to be acoustic fluctuations and treated as background noise. 
However, recent studies ~\cite{gotoda2012chaos, nair2013ijscd, nair2014jfm_multifractality} showed that combustion noise possesses high-dimensional chaotic fluctuations and exhibits multifractal characteristics.
These studies indicate that the combustion noise is dynamically complex,  and should not be considered as noise.
Meanwhile, an intermittency state was also discovered presaging the transition from combustion noise to limit cycle oscillation~\cite{nair2014jfm_intermittency, nair2013chaos}.
When increasing the flow rate in the combustor, intermittent bursts of periodic oscillations with high amplitude can appear randomly among low amplitude aperiodic oscillations. 
Upon increasing the flow rate further, the system will finally enter a regime of periodic oscillation.
The transition is found to be a process in which the system gradually self-organizes and loses its multifractality, as order emerges from chaos~\cite{nair2014jfm_multifractality, unni2015jfm}. 
Several universal power laws were identified~\cite{pavithran2020epl, pavithran2020SR, pavithran2021asme}.
These studies provide valuable early warning signals to evade combustion instability. 

In this study, we attempt to extend our synchronization framework \cite{weng_2020_nody} for the thermoacoustic system further to the turbulent regime and capture the transition from combustion noise to limit cycle oscillation through the route of intermittency.
The model is also expected to identify the self-organization and loss of multifractality in the transition.
Since it is a phenomenological model, the primary goal is to qualitatively replicate the observations from experiments.
%However, it can provide a new configuration for modeling thermoacoustic instability, which considers the nonlinearities of the flame. 
In our previous work \cite{weng_2020_nody}, we showed that a pair of coupled oscillators when coupled with proper coupling strength, can model the quasi-periodic route to chaos observed in laminar thermoacoustic systems. Based on that experience, in this work, we attempt to use a tuned pair of coupled oscillators to represent the simplified turbulent reactive flow. These two oscillators were then, further coupled with a third oscillator representing the acoustic field. We will show that, with proper coupling strength, this three oscillator model can reflect the combustion dynamics arising in a turbulent combustor.
%With proper coupling, these two oscillators are naturally aperiodic when being weakly coupled to a third oscillator, i.e. acoustic oscillator. 

%However, it can provide a new configuration for modeling thermoacoustic instability, which considers the nonlinearities of the flame. 
%With the help of data-driven methods, the model can be further improved and optimized for real combustors.

% check this paper: "Synchronization of two coupled multimode oscillators with time-delayed feedback". Cite it

%Sujith's highly cited paper: Thermoacoustic Instability in a Rijke Tube: Non-Normality and Nonlinearity. Want to talk sth on NL, cite this one

%The reset of the manuscript is organized as follows:
The rest of the manuscript is organized as follows: in Sect.~\ref{Model construction}, we introduce the model based on the synchronization framework; In Sect.~\ref{Results and discussions}, we compare the dynamics observed in the model and the experiment. 
The results will show that the model can qualitatively replicate the transition from combustion noise to periodic oscillation in a turbulent thermoacoustic system. Power laws associated with self organization and loss of multifractality are also captured.

%The history of research on thermocoustic instability
\section{\label{Model construction}Model construction}
Our primary goal is to present a phenomenological, reduced-order model based on a synchronization framework for modeling the transition to thermoacoustic instability in a confined turbulent combustor. 
In our previous study~\cite{weng_2020_nody}, a synchronization-based model was proposed for a laminar combustor, where the acoustic field and the unsteady heat release rate were represented by a pair of nonlinearly coupled damped oscillators.
By varying the coupling strength between the two oscillators, the system can reproduce the route to chaos observed in a laminar combustor~\cite{kabiraj_2012_chaos}.
Multiple bifurcations and different dynamical states from the experiment were captured, including limit cycle oscillation, quasi-periodic oscillation, strange-non chaos~\cite{premraj_2020_EPL}, and chaos.
Similarly, in this work, the acoustic pressure and the unsteady heat release rate are considered as two sub-systems that are nonlinearly coupled. Unlike in a laminar combustor, at the decoupled states, the acoustic and heat release rate fluctuations are disordered due to underlying turbulent fluctuations. This effect needs to be included in the coupled oscillator model. 

From our previous study, we realized that two second-order oscillators with proper coupling can be used to produce chaos~\cite{weng_2020_nody}, and can be used to model the underlying disorder from turbulence. 
Thus, a pair of coupled oscillators $y_1$ and $y_2$ are used to generate the chaotic oscillation that corresponds to combustion noise, when weakly coupled to oscillator $y_3$.
Hence oscillators $y_1$ and $y_2$ form a sub-system representing the simplified turbulent reactive flow, where the time series of $y_2$ represents the unsteady heat release rate. Furthermore, $y_3$ can represent the pressure fluctuations in this three coupled oscillator model.
The three oscillators have natural angular frequencies of $\omega_1$, $\omega_2$, and $\omega_3$. We will use the symbol $f$ to denote the frequency, which is related to angular frequency as $\omega~=~2\pi f$. The damping coefficients for these three oscillators are $\zeta_1$, $\zeta_2$, and $\zeta_3$, respectively.
When decoupled (i.e. the coupling terms are zero), the three oscillators are all damped simple harmonic oscillators, as shown in Eqs. (\ref{eq: oscillator 1}), (\ref{eq: oscillator 2}), and (\ref{eq: oscillator 3}). 

\begin{equation}
\label{eq: oscillator 1}
\ddot y_1+\zeta_{1}\omega_1\dot y_1+\omega_1^2y_1=\rm{NL_1}
\end{equation}

\begin{equation}
\label{eq: oscillator 2}
\ddot y_2+\zeta_2\omega_2\dot y_2+\omega_2^2y_2=\rm{NL_2}
\end{equation}

\begin{equation}
\label{eq: oscillator 3}
\ddot y_3+\zeta_3\omega_3\dot y_3+\omega_3^2y_3=\rm{NL_3}
\end{equation}
where, the terms on the right hand side (NL) are the nonlinear coupling terms. These nonlinear terms used for this model are as follows:
\begin{equation}
\label{eq: NL1}
{\rm NL_1}~=~{C_{12}}(1-{y_2}^2) \dot y_1
\end{equation}

\begin{equation}
\label{eq: NL2}
{\rm NL_2}~=~R_1C_{12}y_1^2+R_2C_{32}y_3^3
\end{equation}

\begin{equation}
\label{eq: NL3}
{\rm NL_3}~=~C_{32}(1-y_2^2)\dot y_3-\dot y_3^3
\end{equation}
Here, we define $C_{12}, C_{21}, C_{23}, C_{32}$ as the coupling strength, where $C_{ij}$ denotes how strongly the $j^{th}$ oscillator affects the dynamics of the $i^{th}$ oscillator. 
Furthermore, the relative strength of the mutual coupling were defined using the ratio of $C_{ij}$s, $R_1~=~C_{21}/C_{12}$, $R_2~=~C_{23}/C_{32}$. 
Recent studies showed that the coupled interaction between the unsteady heat release rate and the acoustic field is asymmetric~\cite{godavarthi_2018_chaos, kurosaka2021chaos}.
Unsteady heat release rate exerts a stronger influence on the acoustic field than the converse, which implies $C_{32}>C_{23}$, or $R_2<1$. 
%The natural frequency of the three oscillators are selected to match the trend of the change in dominant frequency during transition.
%\hl{$\omega_2$ increases linearly from $\pi$ to $2\pi$ as $C_{32}$ varies from 10 to 20, which gives a natural frequency $f_2~=~1$ in the regime of periodic oscillation.} %The parameter values used in the Eq: {\ref{eq: oscillator 1}}-{\ref{eq: NL3}} are: $\omega_1~=~1.125 \omega_2$, $\omega_2~=~1$, $\omega_3~=~1.3$, $\zeta_{1}~=~0.1$, $\zeta_{2}~=~0.05$, $\zeta_{3}~=~0.3$, $C_{12}~=~1.8$, $R_1~=~1.2$, and $R_2~=~0.54$.
We select the natural frequencies of the three oscillators in a way that the output time series can match the trend of the dominant frequency in the experiment.
To achieve this, we assign a fixed ratio between the natural frequencies of oscillator $y_1$ and $y_2$ ($\omega_1~=~1.125 \omega_2$), resulting in intended aperiodic oscillation when weakly coupled to oscillator $y_3$.
Meanwhile, to ensure that the dominant frequency of $\dot q'$ increases during the transition as observed in the experiment, we set a linear relation between the natural frequency, $\omega_2$ and the coupling strength, $C_{32}$ ($\omega_2~=~2\pi C_{32}/20$). The natural frequency of the acoustics ($y_3$) is chosen to be constant ($\omega_3~=~1.3\times 2\pi$), since in the experiments the natural frequency of the combustor geometry is fixed. 
%$y_3$ is chosen with a fixed natural frequency, where $\omega_3~=~1.3\times 2\pi$.
The other parameter values used in the Eq.~({\ref{eq: oscillator 1}}-{\ref{eq: NL3}}) are selected to be: $\zeta_{1}~=~0.1$, $\zeta_{2}~=~0.05$, $\zeta_{3}~=~0.3$, $C_{12}~=~1.62$, $R_1~=~1.2$, and $R_2~=~0.54$.
In the study, since $y_1$ and $y_2$ form a subsystem that represents unsteady heat release rate, $C_{12}$ is fixed. Meanwhile, $C_{32}$ is considered the control parameter which dictates the transitions between different dynamical states of the system. In an experimental turbulent thermoacoustic system, the control parameter is generally the Reynolds number of the flow, $Re$=$\bar{u}L/\nu$, where $L$ is the characteristic length scale of the burner and $\nu$ is the kinematic viscosity of the unburned mixture.
With the current model and the value of parameters mentioned above, the transition between various states was observed for the range of $C_{32}$ from 14 to 18. The coupled equations (Eq: \ref{eq: oscillator 1}-\ref{eq: oscillator 3}) are solved using the 4th order Runge-Kutta method in Matlab. Gaussian noise with variance $\sigma=5\%$ is then added to the time series.
We note that for a fixed $C_{32}$, the model outputs normalized time series of heat release rate and pressure fluctuations.

As mentioned earlier the primary goal of this work is to demonstrate that coupled oscillator models can reproduce combustion dynamics in a turbulent combustor. During construction of the model we intended to keep the nonlinear terms as simple (low-order) as possible while qualitatively reproducing the key dynamics observed in experiments. The current forms of nonlinear terms shown in Eq: \ref{eq: NL1}-\ref{eq: NL3} and the values of various constants were chosen after manual optimization and satisfy the requirements. We note that, in general, a large number of possible forms of nonlinear terms can be selected. A robust method for identifying and optimizing the nonlinear terms will be focus of our future work.

%\hl{address the added noise}

%It is to be noted that for a fixed $C_{32}$, the model outputs normalized time series of heat release rate and pressure fluctuations. In an experimental turbulent thermoacoustic system, the control parameter is generally the flow rate of the Reynolds number or $Re$\hl{=$\bar{u}d/\nu$, where $d$ is the burner diameter and $\nu$ is the kinematic viscosity of the unburned mixture. (The definition to be verified)}  

%Note that, the parameters and nonlinear terms in the model are manually selected.
% Other combination of nonlinear terms may also provide the same route of transition, and nonlinear terms with higher-order or more complex combinations can show a better result matching the experiment. 
% However, we try to choose the simplest terms that can capture the experimental observation qualitatively. 
% Because we intend to show that it is possible to build a reduce-order model for turbulent thermoacoustic systems based on the synchronization framework and give its basic configuration.
% This configuration will be useful to develop models for engineering applications by introducing data-driven assessments of model parameters, in the future.

\section{Experimental Details}
We primarily use the experimental data from Pawar et al.~\cite{pawar_2017_jfm} as a benchmark to validate the model.
The experimental study applies to a laboratory-scale turbulent combustor with a bluff body stabilized flame, where liquefied petroleum gas was burned with air. The incoming airflow first passes through a settling chamber. Then, it is premixed with the fuel and sent into the combustion section. The pressure fluctuation ($p'$) in the combustor is estimated using pressure transducers. 
The heat release rate ($\dot q$) is acquired by measuring the chemiluminescence intensity of CH* using a photo-multiplier tube (PMT), from which its fluctuations ($\dot q'$) is calculated. 
%Furthermore, in a fully developed turbulence, turbulence intensity, $u_{rms}/\bar{u}$, and hence, heat release intensity $u_{rms}/\bar{u}$ is expected to be constant with $Re$. Thus, it is appropriate to use heat release rate intensity $u_{rms}/\bar{u}$
%Naturally, when coupled with acoustics in the experiments in closed combustor, the fluctuation in heat release ($\dot q'$) will be affected by both $\bar{q}$ and acoustics. Since, from the modeling perspective, we are only interested in assessing the effect of acoustics on heat release rate fluctuations, it is appropriate to use  
%although the total enthalpy (or energy) of the system which changes the characteristic frequencies. Thus, to compare the time series of the model output with the experimental data (shown later), the former was dimensionalized by multiplying with appropriate $\omega$. It is worth noting that for an experiment with laminar combustor, the transition between dynamical states where achieved generally by moving the flame location keeping the flow rate constant, and such re-normalization was not required in our previous work with laminar system \cite{weng_2020_nody}.
During the experiment, the fuel flow rate was kept constant at 25 slpm, while the flow rate of air varied from 400 slpm to 940 slpm, covering Reynolds number ($Re$) range from $1.09\times10^5$ to $2.12\times10^5$. We note that an increase in flow rate ($Re$) increases the mean heat release rate ($\bar{q}$) of an unconfined (de-coupled from acoustics) turbulent flame due to decreased residence time \cite{peters2000turbulent}. Thus, it is appropriate to use heat release rate intensity $\dot q'_{rms}/\bar{\dot q}$ as the statistical measure of fluctuations. The acoustic characteristics of the chamber, however, remain unaffected by the $Re$, thus $p'_{rms}$ is a good measure. 

From the measurements, we confirm that the combustor exhibited stable operation at low $Re$. However, due to the inherent turbulence, the pressure and heat release rate keep fluctuating at a low amplitude. This regime refers to high dimensional chaos~\cite{tony2015PRE}. As $Re$ increases, the amplitude of pressure oscillation increases continuously. Meanwhile, among aperiodic oscillations, burst and periodic oscillations start emerging in a near-random fashion, which refers to the state of intermittency. When $Re$ further increases, both pressure and heat release rate oscillate periodically, and with significantly greater amplitude, corresponding to the onset of thermoacoustic instability.

We will use a second set of experimental data from Nair et al.~\cite{nair2014jfm_intermittency} during our discussion on the power-law behaviors in the transition process (Sec: \ref{sec: power law}). This work used a swirl stabilized burner. The rest of the configuration and experimental technique remained mostly unchanged. The details of this setup can be found in~\cite{nair2014jfm_intermittency}.

{%As the flow rate increases, the system approaches the onset of thermoacoustic instability, and both $p'$ and $\dot q'$ grow. 
In experiments, the flow in the combustor is fully developed (constant turbulence intensity, $u'/\bar u$) and hence, $u'$ grows linearly with $Re$ ($u' \propto Re$).
In these experiments, the measured $p'$ and $\dot q'$ which increase with $Re$ are composed of the contribution from turbulence and the thermoacoustic instability. The former arises from the fluctuations in flow velocity $u'$, and hence can be assumed to be proportional to $Re$. These turbulence-induced effects are apparent when one measures the pressure in a turbulent flow inside the combustor without the flame or heat release rate in an unconfined flame, thus no thermoacoustic effects. It is to be noted that the model presented in Eq: {\ref{eq: oscillator 1}} - {\ref{eq: NL3}}, only accounts for the coupled dynamics between acoustics and heat release, thus discounts the increase in $p'$ and $\dot q'$ due to turbulence effects. Since, these discounted effects are proportional to $Re$, and in our model, we assumed $C_{32}\sim Re$, for comparison purposes, the experiment equivalent of pressure and heat release rate from our models can be obtained as $p' = y_3C_{32}$, and $\dot q' =  y_2C_{32}$.}

% Similarly, the measured $\dot q'$ 
% For the pressure fluctuations $p'$, while a part of growth comes from thermoacoustic instability, the increasing $u'$ also introduces a higher magnitude of $p'$, even when there is no flame. The part of $p'$ caused by increased $u'$ should be canceled by normalizing $p'$ by $Re$.
% For the heat release rate fluctuation $\dot q'$, as $u'$ increases, turbulence introduces more wrinkles to the local flame surface. 
% The augmented flame surface area results in a higher mean heat release rate $\bar{\dot{q}}$, irrespective of thermoacoustic instability. So $\dot q'$ needs to be normalized by $\bar{\dot{q}}$, where $\bar{\dot{q}} \propto Re$.
% Thus, we consider $y_5$ and $y_3$ to represent $p'$ and $\dot q'$ that are normalized by $Re$. 
% We assume $C_{32} \propto Re$, so we have $y_3C_{32} \sim p'$, and $y_2C_{32} \sim \dot q'$.}

\section{\label{Results and discussions}Results and discussions}
The presented synchronization based three-oscillator model is expected to replicate some of the key features of a turbulent combustor. The model should capture the transition from combustion noise to periodic oscillation, during which the amplitude of pressure and heat release rate oscillation increase monotonically. It should also predict different synchronous states during the transition, which mostly depends on the synchronization states of the pressure and heat release rate.
Moreover, in the regime of combustion noise, the model is expected produce a multifractal behavior in pressure oscillations, which is the hallmark of a stable turbulent combustor. As the system transitions to thermoacoustic instability, self-organization is observed during experiments~\cite{pavithran2020epl, pavithran2020SR} and the time series shows a scale-independent nature marked by power-law dynamics. We expect the model to be able to predict similar behavior.

In the following subsections, we will analyze the statistics and dynamics of the time series produced by the model, and qualitatively compare with those of the experimental data. We note that the primary expectation from the model with predefined model parameters is to qualitatively reproduce the experimental observations. The quantitative similarity is not expected, which requires tuning the model parameters based on the experiments.

\begin{figure}[ht!]
\centering
\includegraphics[width=0.7\textwidth]{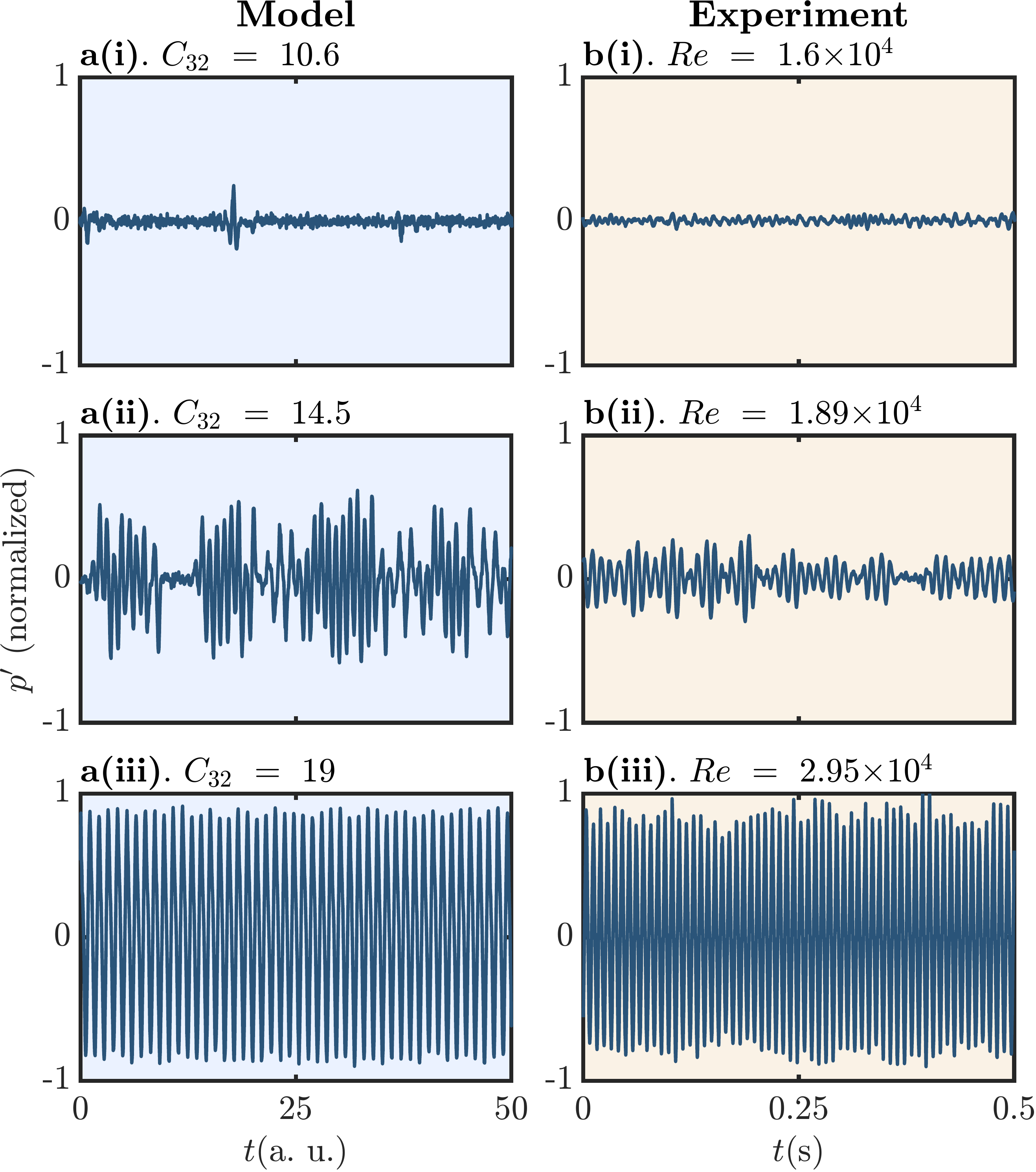}
\caption {Time series obtained from the model and experiments at different dynamical states. From left to right: model, experimental results from Pawar et al. \cite{pawar_2017_jfm} From top to bottom: combustion noise (\textbf{a(i)}: $C_{32}~=~11.05$, \textbf{b(i)}: $Re~=~1.6\times10^4$), intermittency (\textbf{a(ii)}: $C_{32}~=~14.95$, \textbf{b(ii)}: $Re~=~1.89\times10^4$), limit cycle oscillation (\textbf{a(iii)}: $C_{32}~=~19.7$, \textbf{a(iii)}: $Re~=~2.95\times10^4$)}
\label{fig: 1}
\end{figure}

\subsection{The temporal transition from combustion noise to periodic oscillation}
First, we will use the temporal dynamics and their statistics in assessing the model performance. 
As mentioned before, in the model, the control parameter is the coupling strength between unsteady heat release rate and the acoustic oscillator $C_{32}$. By increasing the coupling strength $C_{32}$, we observe that the system transitions from chaos to periodic limit cycle oscillation through intermittency, a behavior also observed in experiments. Figure.~\ref{fig: 1}a(i, ii, iii) shows the representative time series of the normalized pressure functions from the model at different $C_{32}$, which show the changes in morphology of the time series at different dynamical states. Figure.~\ref{fig: 1}b(i, ii, iii) shows a similar time series obtained from experiments (\cite{pawar_2017_jfm}) at different $Re$, which clearly exhibit very similar qualitative behaviors.
%\hl{are from experiments. Both experiments are turbulent combustors, Pawar's experiment uses a bluff body as the flame stabilizer, while Nair uses a fixed vane swirler. }
As shown in Fig.~\ref{fig: 1}a(i), at low $C_{32}$, the signal is disordered and the overall amplitude of $p'$ is small, which corresponds to the regime of combustion noise.
In Fig.~\ref{fig: 1}a(ii), as $C_{32}$ increases, although the time series remains disordered, the amplitude of the time series increases, and intermittent oscillations appear. 
During intermittency, the signals are composed of aperiodic, low-amplitude fluctuations interspersed amongst periodic, high amplitude fluctuations in a near-random fashion. The different parts of the signal have amplitudes with several different orders of magnitudes~\cite{nair2014jfm_intermittency}. 
Previous studies proposed a route from combustion noise to thermoacoustic instability through intermittency~\cite{nair2014jfm_intermittency}, and hence it is important to determine the existence of intermittency state in the model.
When we further increase $C_{32}$, as shown in Fig.~\ref{fig: 1}a(iii) the time series becomes more regular and with a larger amplitude, the periodic oscillation has now become dominant. It indicates that the system has entered the regime of limit cycle oscillation. As mentioned earlier, we observe similar characteristics in the corresponding time series from the experiment, shown in columns Fig.~\ref{fig: 1}b(i, ii, iii). 

\subsection{The variation in the \textit{RMS} and dominant frequency of $p'$ and $\dot{q'}$}

\begin{figure}[ht]
\centering
\includegraphics[width=0.9\textwidth]{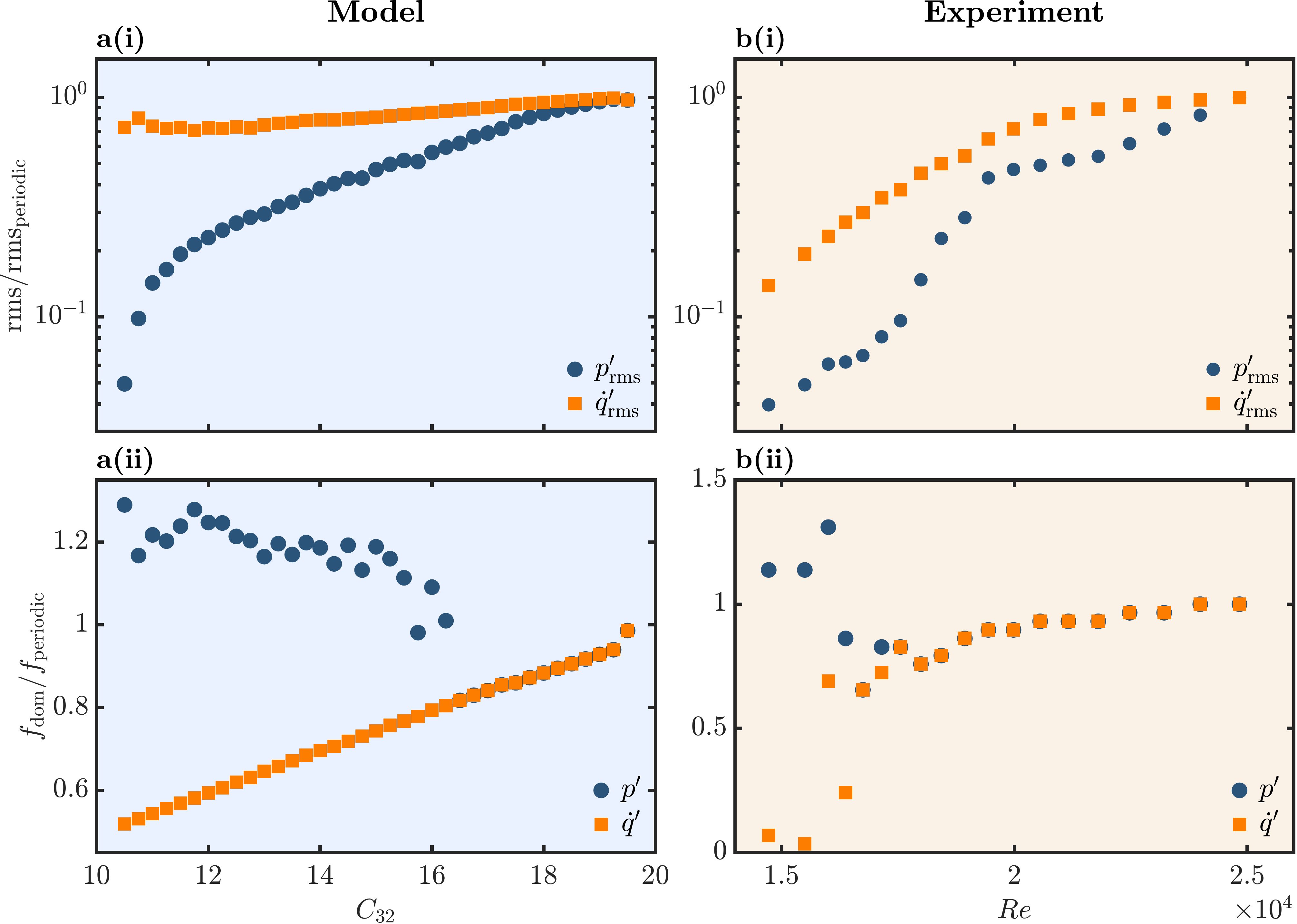}
\caption {\textbf{a(i)} and \textbf{b(i)}: $RMS$ of $p'$ and $\dot q'$ at different conditions from model and experiment.
As $C_{32}$ or $Re$ increases, $p'_{\rm rms}$ increases monotonically. 
\textbf{a(ii)} and \textbf{b(ii)}: dominant frequency of $p'$ and $\dot q'$ at different conditions from model and experiment.
The dominant frequency of pressure and heat release rate when $C_{32}$ or $Re$ varies.
As $C_{32}$ or $Re$ increases and the system transitions to periodic oscillation, both signals start to have the same dominant frequency.}
\label{fig: rms and fd}
\end{figure}

One of the hallmarks of thermoacoustic instability is the large increase in amplitude of 
pressure and heat release rate fluctuations as the system transitions from combustion noise to the limit cycle oscillation. While this can be visualized qualitatively for three different $C_{32}$ (model) and $Re$ (experiment) in Fig. \ref{fig: 1}, we additionally compare the root mean square values of both pressure fluctuations ($p'_{\rm rms}$) and heat release rate fluctuations ($q'_{\rm rms}$) for a range of operating conditions (Fig.~\ref{fig: rms and fd}a(i) and b(ii)). Here we evaluate the $\textit{RMS}$ of time series of a quantity, $x(t)$ as $x_{\rm rms}~=~\sqrt{\frac{1}{M}\sum_{i=1}^{M} (x(t_i)-\bar{x})}$, where $\bar{x}$ is the mean of $x$ and $M$ is the length of the time series. Our model predicts a continuous increase of $p'_{\rm rms}$ and $q'_{\rm rms}$ with increase in the control variable, $C_{32}$ (Fig. \ref{fig: rms and fd}a(i)), a trend which is observed for experiments with turbulent combustor (Fig. \ref{fig: rms and fd}b(i)). 

% (a1, b1) shows the variation of the root mean square of pressure time series ($p'_{\rm rms}$) when changing the coupling strength $C_{32}$. $p'_{\rm rms}$ here is calculated as Eqn.~\ref{eq: prms}
% \begin{equation}
% \label{eq: prms}
% p'_{\rm rms}~=~\sqrt{\frac{1}{M}\sum_{i=1}^{M} p'(t_i)}
% \end{equation}

%The corresponding time series from the experiment is shown in columns Fig.~\ref{fig: 1}(b). The model present the similar transition when increasing the coupling strength comparing to the experiments.

\begin{figure}[ht]
\centering
\includegraphics[width=0.95\textwidth]{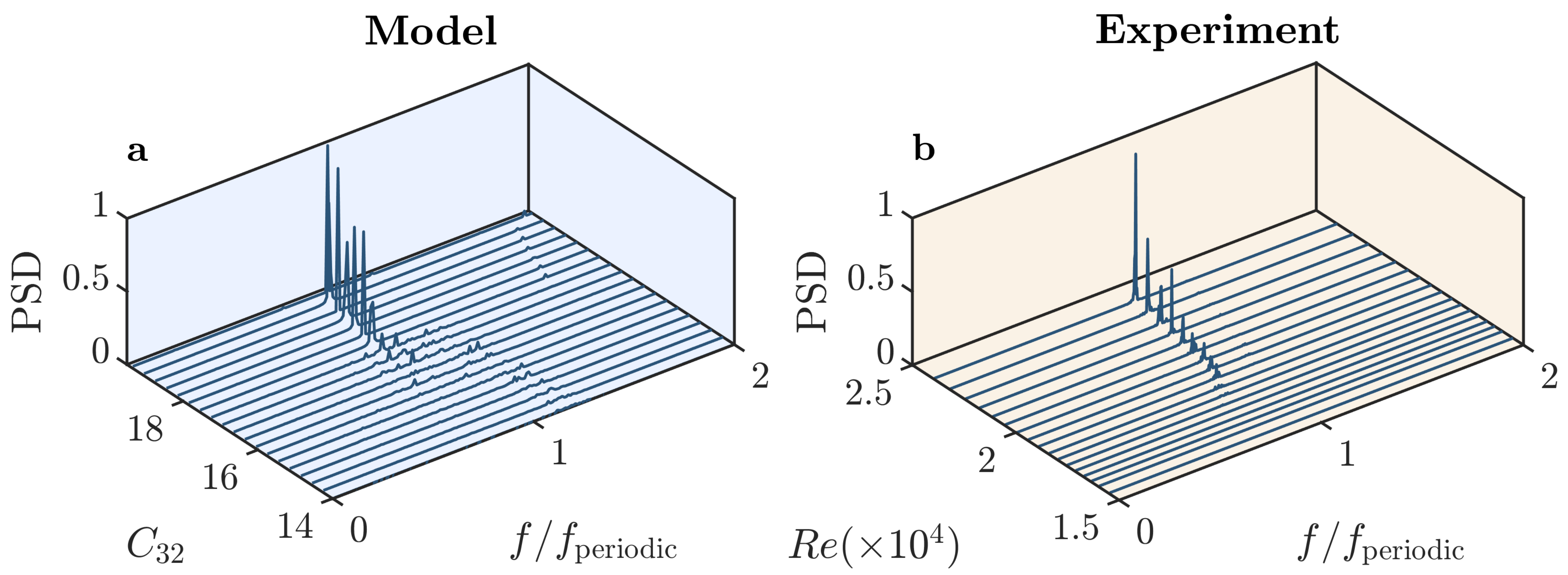}
\caption {The power spectra density of $p'$ at different conditions from model and experiment. 
%For better visualization, the spectra are plotted with a resolution of 4Hz. 
As $C_{32}$ or $Re$ increases, the height of the peak increases significantly, where periodic oscillation becomes dominant.}
\label{fig: water_fall}
\end{figure}

We next analyze the dominant frequency of the pressure and heat release rate signals, which is critical in determining the mode of oscillations in thermoacoustic instabilities. Figure~\ref{fig: rms and fd}a(ii) depicts the variations of dominant frequencies of $p'$ and $\dot q'$ with $C_{32}$, predicted by the model. 
As we can see in the figure, initially at low $C_{32}$ during the state of combustion noise, the two signals exhibit different dominant frequencies, which are far apart. But as $C_{32}$ increases, the two dominant frequencies gradually merge, and finally, both signals have the same frequency during limit cycle oscillation (high $C_{32}$). 
This suggests that the two signals are synchronized with each other (we will quantify this process in the next subsection). A similar behavior of dominant frequency is also observed for experiments as $Re$ is increased (Fig. \ref{fig: rms and fd}b(ii)).
%the dominant frequency may jump back and forth in the transition zone. In real systems, the transition may not strictly go through this route.
Furthermore, we also compare the power spectral density (PSD) of $p'$ at various $C_{32}$ (model) and $Re$ (experiment) in Fig. \ref{fig: water_fall}. They indeed show a similar trend in that, the peak of the dominant frequency becomes sharper in the spectrum as $C_{32}$ or $Re$ increases, and periodic oscillations become more dominant (seen in Fig. \ref{fig: 1}). 
The sharpened peak indicates that the energy is gradually concentrated to the dominant frequency from a previously broadband spectrum at low $C_{32}$ or $Re$. We will quantify and discuss the behavior of spectral condensation in the later section.

\begin{figure}[hb!]
\centering
\includegraphics[width=0.85\textwidth]{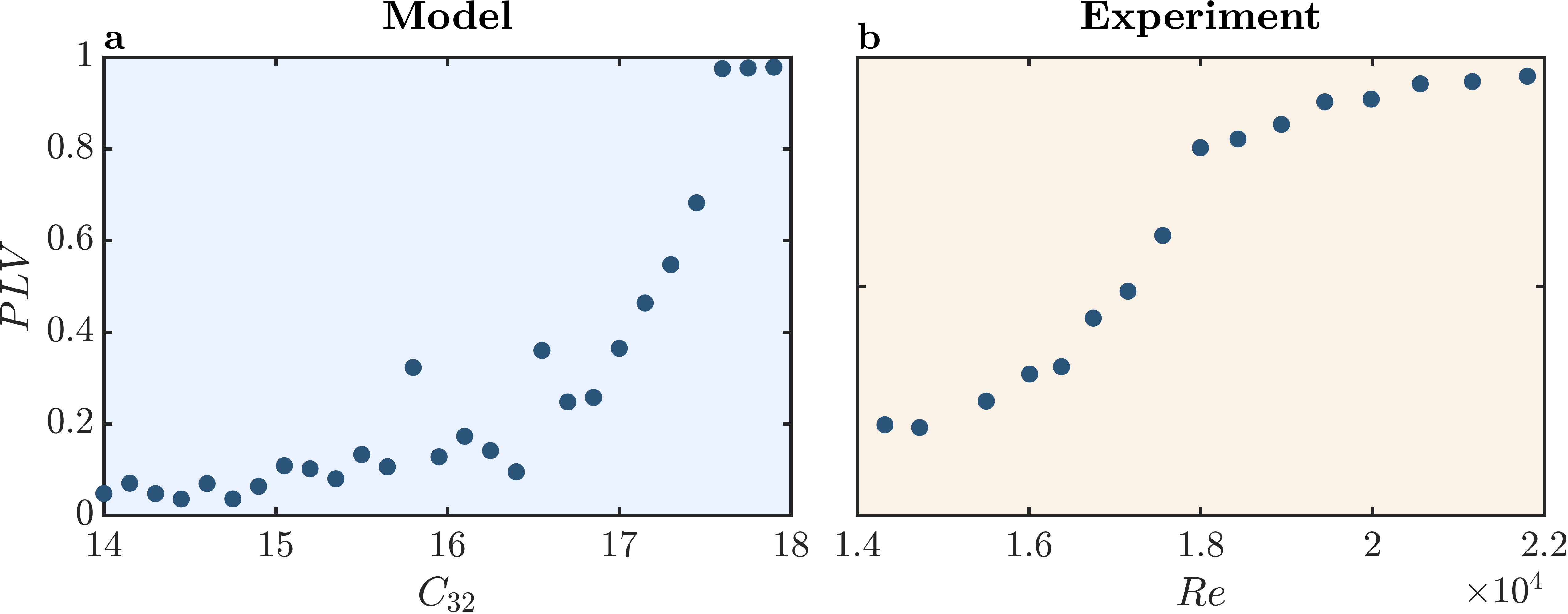}
\caption {$PLV$ between $p'$ and $\dot q'$ at different conditions from model (\textbf{a}) and experiment (\textbf{b}). In the regime of combustion noise, the phase lock value between $p'$ And $\dot q'$ is close to 0, which indicates the two signals are desynchronized. As periodic oscillation appears, $PLV$ increases and stays close to 1, which corresponds to synchronization.}
\label{fig: PLV}
\end{figure}

\subsection{Synchronization characteristics}
As mentioned in the introduction, experimental studies have shown that the time series of $p'$ and $\dot q'$ exhibit a de-synchronous state during the state of combustion noise, in that the phase of their oscillation varies with time. The studies have also confirmed that these time series become phase synchronized as the system approaches the onset of combustion (thermoacoustic) instability. During limit cycle oscillation, $p'$ and $\dot q'$ become completely phase locked or phase synchronized. The degree of synchronization between between $p'$ and $\dot q'$ can be evaluated with the \textit{Phase Locking Value} ($PLV$), %. It helps us detect the different status of coupling between $p$ and $\dot q'$.
which can be calculated using the Hilbert transform and the relation~\cite{panter1965book}, 
\begin{equation}
\label{eq: PLV}
PLV~=~\frac{1}{M}\left|\sum_{i=1}^{M} e^{j\left(\phi_{p'}\left(t\right)-\phi_{\dot q'}\left(t\right)\right)}\right|,
\end{equation}
where $\phi_{p'}(t)$ and $\phi_{\dot q'}(t)$ are the instantaneous phase of the pressure and heat release rate fluctuations. PLV can vary between 0 to 1; a $PLV$ value of 0 indicates that the two signals are desynchronized, and a $PLV$ value of 1 corresponds to perfect phase synchronization. The $PLV$ evaluated based on the experimental data \cite{pawar_2017_jfm} plotted in Fig.~\ref{fig: PLV}(b), shows low values of $PLV$ at low $Re$ (control parameter for the experiments) and at high $Re$, $PLV$ approaches 1, suggesting phase synchronization. To assess if our model can predict such characteristics, we calculated the $PLV$ from the model output and then plot as function of the coupling strength, $C_{32}$ in Fig.~\ref{fig: PLV}(a). The time series of $p'$ and $\dot q'$ are indeed de-synchronized during at a low $C_{32}$ (combustion noise). Similar to experiments, the model also predicts an increase in $PLV$ with $C_{32}$ (control parameter for the model). Finally $PLV$ becomes almost 1 when $C_{32}>17$, suggesting phase synchronization during limit cycle oscillations. 
% As expected, the regime of desynchronization where the phase lock value is close to 0 corresponds to combustion noise. 
% As $C_{32}$ increases and the system transitions to limit cycle oscillation, the phase lock value increases to 1, indicating that the two signals are phase synchronized. These observations are consistent with the experimental study~\cite{pawar_2017_jfm}.
%\hl{may need to distinguish phase sync and GS here}

\subsection{Loss of multifractality during transition}
\label{subsection: multifractal}
In a turbulent combustor, the observed flame dynamics is an outcome of continuous interaction of multi-scale eddies in the flow with the chemical reaction. Such interaction leads to a broadband low amplitude oscillation in the heat release rate and pressure fluctuations, even when they are in the de-synchronized states. 
Such broadband oscillations or combustion noise has been traditionally assumed to be stochastic and has been modeled as white or colored noise in many studies~\cite{noiray2017JEGTP, tony2015PRE, clavin1994CST, zhang2021nody}. 
%\hl{ADD References}. 
%In turbulent combustors, even when the burner operates at stable conditions, because of turbulence, low amplitude irregular pressure fluctuation always exists. The fluctuation is known as combustion noise, which is generated by the turbulent reactive flow and is with broadband characteristics. Traditionally, it is assumed to be stochastic and usually modeled as white noise in many studies.
However, recent studies have shown the combustion noise is not always stochastic, and it has features of higher-dimensional chaos~\cite{gotoda2012chaos, gotoda2015pre}. Furthermore, experimental data and analyses have also shown that the pressure fluctuations from combustion noise can be self-similar and exhibit multifractality~\cite{nair2013ijscd, nair2014jfm_multifractality, unni2015jfm}. 
Such behavior of multifractality is, indeed, a hallmark of turbulent flows~\cite{sreenivasan1991fractals, meneveau1987simple, prasad1988multifractal}  and turbulent flame dynamics~\cite{raghunathan2020jfm}. %~\cite{saha2014pof, unni2018chaos}. 
In other words, combustion noise caused by turbulent reactive flows is dynamically complex and can be considered deterministic. 
More recent studies showed that the onset of thermoacoustic instability can be characterized as an order emerging from chaos via a state of intermittency~\cite{juniper2018AR}. 
During this transition, the system gradually loses its multifractality~\cite{nair2014jfm_multifractality, unni2015jfm}. Based on these observations, several techniques are proposed to determine the stability boundaries and to provide early warning signals (EWS)~\cite{nair2014jfm_multifractality}. 
Since the proposed model aims to replicate the dynamics of turbulent combustors, it is critical to assess if it can reproduce the presence and loss of multifractality in pressure fluctuations in combustion noise and thermoacoustic instability (limit cycle oscillations), respectively. 

%\hl{talk about multifractality}
In time series analysis, the fractal dimensions ($D$) are often used to characterize fractal behavior, where the value and the length of the data show a self-similarity associated with a power law, and $D$ is the exponent of the scaling relation. 
Fractal nature also exists in many complex systems, e.g. time signals with complex dynamics can be self-similar in local fluctuations with multiple scales~\cite{mandelbrot1982book}. %\hl{REFERENCE}. 
In a self-similar time series, the fractal dimension is related to the Hurst exponent $(H)$ as $D=2-H$~\cite{west1994PR_fractal}, which evaluates the long-term memory of the time series. 
However, in many systems involving broadband perturbations (e.g. turbulence), the scaling behavior can be more complex and may depend on the order of the structure function, giving rise to multifractality. 
A multifractal system, instead of one single Hurst exponent, requires a singularity spectrum consisting of a series of generalized Hurst exponents with multiple orders to describe its characteristics~\cite{harte2001book_multifractals}. The singularity spectrum depicts the fractal behaviors of local fluctuations of different magnitudes.
In our work, the generalized Hurst exponents are evaluated by multifractal spectrum $f(\alpha)$ versus singularity exponent $\alpha$ using the multifractal detrended fluctuation analysis (MFDFA)~\cite{kantelhardt2002PA, ihlen2012FIP}.In Appendix, we provide a brief description of MFDFA.

\begin{figure}[ht] 
\centering
\includegraphics[width=0.85\textwidth]{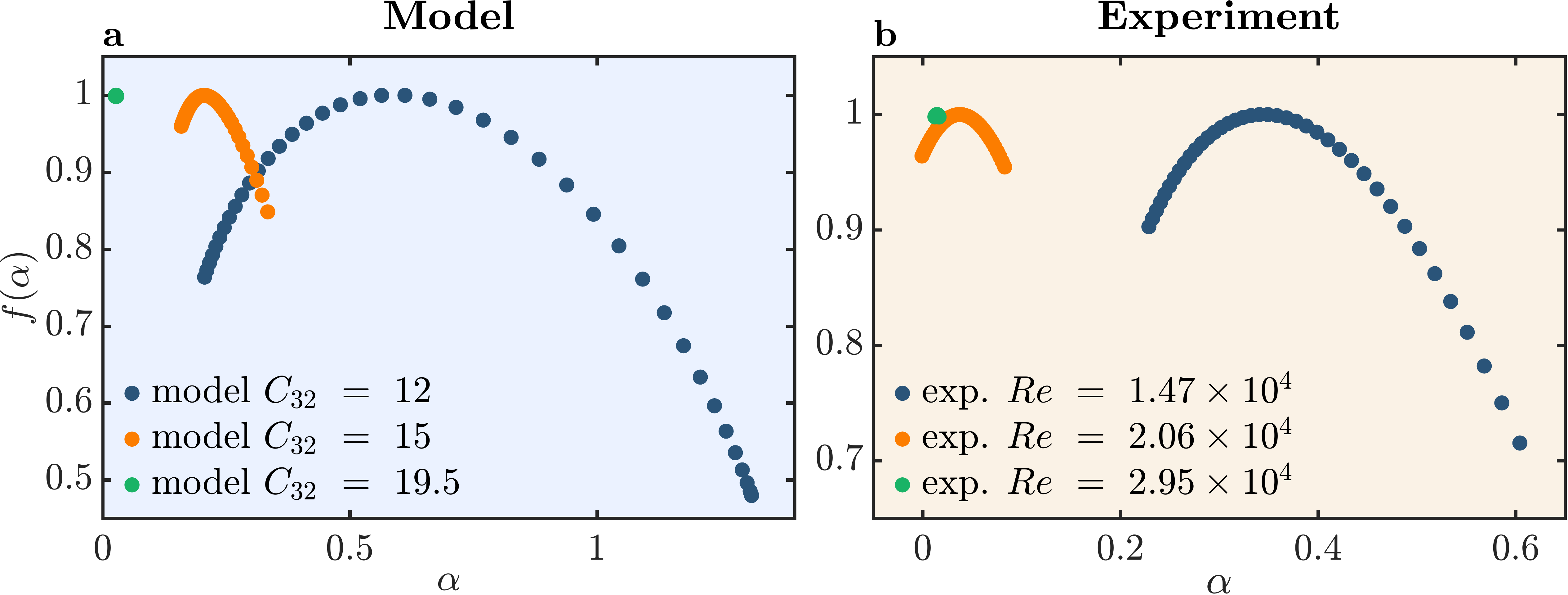}
\caption {Multifractal spectrum at different dynamical states from the model (\textbf{a}) and experiment (\textbf{b}), as $C_{32}$ or $Re$ increases, the system transitions from combustion noise to periodic oscillation. In the regime of combustion noise, the multifractal profile exhibits a wide range spectrum, which corresponds to a dynamically complex state. During the transition, the spectrum shrinks, indicating that the system gradually loses its multifractality.}
\label{fig: 5}
\end{figure}

Using MFDFA, the multifractal spectra $f(\alpha)$ for pressure oscillations obtained from our model and previous experiment \cite{pawar_2017_jfm} are plotted as function of $\alpha$ in Fig.~\ref{fig: 5}a and b, respectively. It is to be noted that the width of the spectrum quantifies the degree of multifractality of the time series. Figure \ref{fig: 5}a shows that at low $C_{32}$ (combustion noise) the spectrum is indeed wide, suggesting strong multifractality and hence complexity. This behavior is also observed for experimental data at low $Re$ (Fig. \ref{fig: 5}b).
% In our study, as shown in Fig.~\ref{fig: 5}, when the coupling strength is low, where the system is in the regime of combustion noise, we observe a wide multifractal spectrum.
% It indicates the pressure signal in this regime is dynamically complex, and exhibit multifractality. 
As $C_{32}$ increases, the width of the spectrum shrinks, indicating that the system gradually loses its multifractality as it approaches thermoacoustic instability. At high $C_{32}$, when the system exhibits limit cycle oscillations, the spectra condenses to a point, and hence the multifractality has been lost. This reduction of width and the collapse of the multifractal spectrum, indeed, can also been seen in experimental data as $Re$ is increased (Fig. \ref{fig: 5}b).  

% in experimental observations~\cite{nair2014jfm_multifractality, unni2015jfm}.

\subsection{Power law in self-organization} \label{sec: power law}
In this section, we will discuss the recent discoveries of scaling relations observed as a turbulent system transitions to thermoacoustic instabilities, and assess if the proposed coupled oscillator model can also produce such dynamics. Turbulent systems generally contain high-dimensional chaos, and their time series is naturally irregular and disordered. However, when turbulence is coupled to the other subsystems, through their inherent coupling, a self-organization may occur, in that the underlying broadband nature is lost. One typical example is the turbulent Rayleigh-Bénard system, where the fluid motion is turbulent at a low-temperature gradient.
However, at a high-temperature gradient, fluid particles may move orderly and result in large-scale spatial patterns~\cite{pandey2018nature}.
During self-organization, the positive feedback leads the disordered, chaotic system to transition to an ordered state, and regularly repeating patterns emerge both time-wise and space-wise.
The initially disordered state is generally a superposition of a series of motions with a wide range of modes and a broadband energy spectrum. 
During self-organization, one or more modes gradually emerge and become dominant.
Self-organization has been studied extensively in various turbulent systems including aeroelastic~\cite{hansen2007wind} and aeroacoustic~\cite{flandro2003aiaa} systems.
In recent years, studies found that the transition from chaotic fluctuations to periodic oscillations in the turbulent thermoacoustic system is also a process of self-organization~\cite{unni2015jfm}. Several self-similar behaviors were subsequently identified, including the power laws of Hurst exponent~\cite{pavithran2020epl} and spectral condensation~\cite{pavithran2020SR}.
To validate the proposed model, we will assess the scaling laws of the Hurst exponents and spectral condensation using the methodology reported by Pavithran et al.~\cite{pavithran2020SR, pavithran2020epl}. 
Simultaneously we will use the same measures on two sets of different experiments, bluff-body stabilized~\cite{pawar_2017_jfm} and swirl-stabilized~\cite{nair2014jfm_intermittency} combustors. 

\begin{figure}[ht]
\centering
\includegraphics[width=0.65\textwidth]{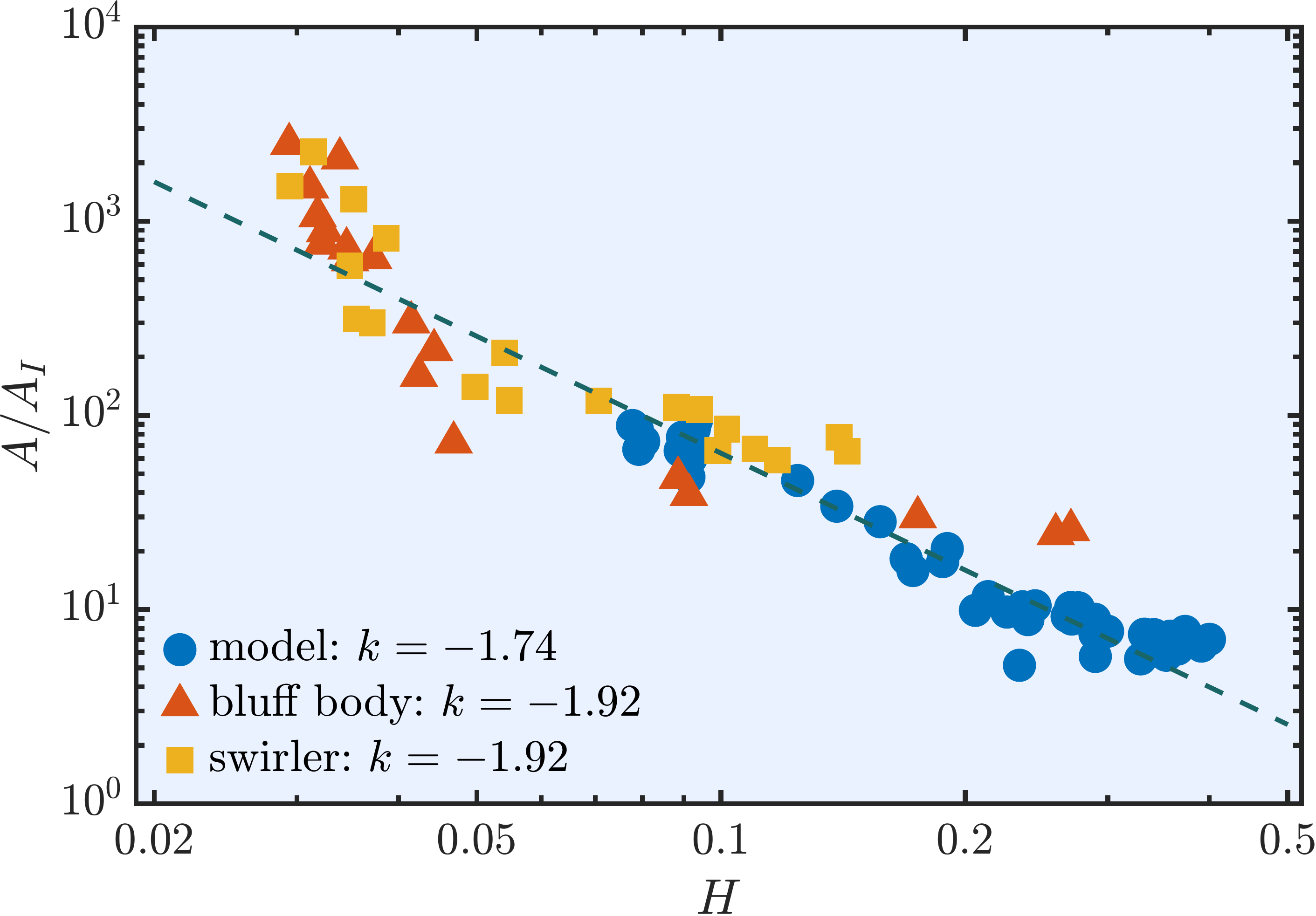}
\caption {The spectral amplitude of the dominant mode ($A$) versus Hurst exponents ($H$) at different control parameters during the transition to thermoacoustic instability.
Note that both $A$ and $H$ are plotted on a logarithmic scale, and $A$ is normalized by the $A_I$, which is the $y$-intercept of linear regression: $\log (A)~=~n \log (H) + \log (A_I)$.
We observed that an inverse power law, $A/A_I\propto H^{n}$ with an exponent $n=-2$ (dashed line) fits the data from both model and the two experimental studies. Experimental data are from bluff-body-stabilized combustor by Pawar et al. \cite{pawar_2017_jfm}, and swirl-flow combustor by Nair et al.~\cite{nair2014jfm_intermittency}.
}
\label{fig: 8}
\end{figure}

%The Hurst exponent ($H$) quantifies the long-range dependence of a time series by evaluating the local trend of fluctuation.
As discussed in \ref{subsection: multifractal}, the Hurst exponent is related to fractal dimension and can vary between 0 and 1. 
Time series with $H~=~0.5$ refers to uncorrelated random fluctuations. 
For upper range ($0.5<H<1$), the time series is persistent with a long-term positive auto-correlation, i. e. a high value in the time series is likely to be followed by another high value. On the other hand, for $0<H<0.5$, the time series is antipersisitent, i. e. it tends to switch between high and low values.
In recent studies ~\cite{pavithran2020SR, pavithran2021asme}, an inverse power-law relation between the spectral amplitude of the dominant mode ($A$) and the Hurst exponent ($H \propto A^n$) was reported in turbulent systems that exhibits transition from chaotic state to periodic state and thus, self-organization. Interestingly, the exponent $n$ was found to be almost constant and $n\approx-2$ for a series of different systems, including aeroacoustic, aeroelastic, and thermoacoustic systems.
Here, we calculated the Hurst exponent and spectral amplitude of the dominant mode during the transition from the data produced by the model. 
The results are compared with the experimental data from the bluff-body and swirler combustors, as shown in Fig.~\ref{fig: 8}. 
We, indeed, observe a power law with a exponent close to -2 for both the model and the experiments. Such agreement implies that the model is capable in capturing the underlying path in the transition to thermoacoutsic instability in turbulent combustors. 
%The path is universal regardless of the configurations of the combustor, and the model can capture it.

\begin{figure}[ht]
\centering
\includegraphics[width=1\textwidth]{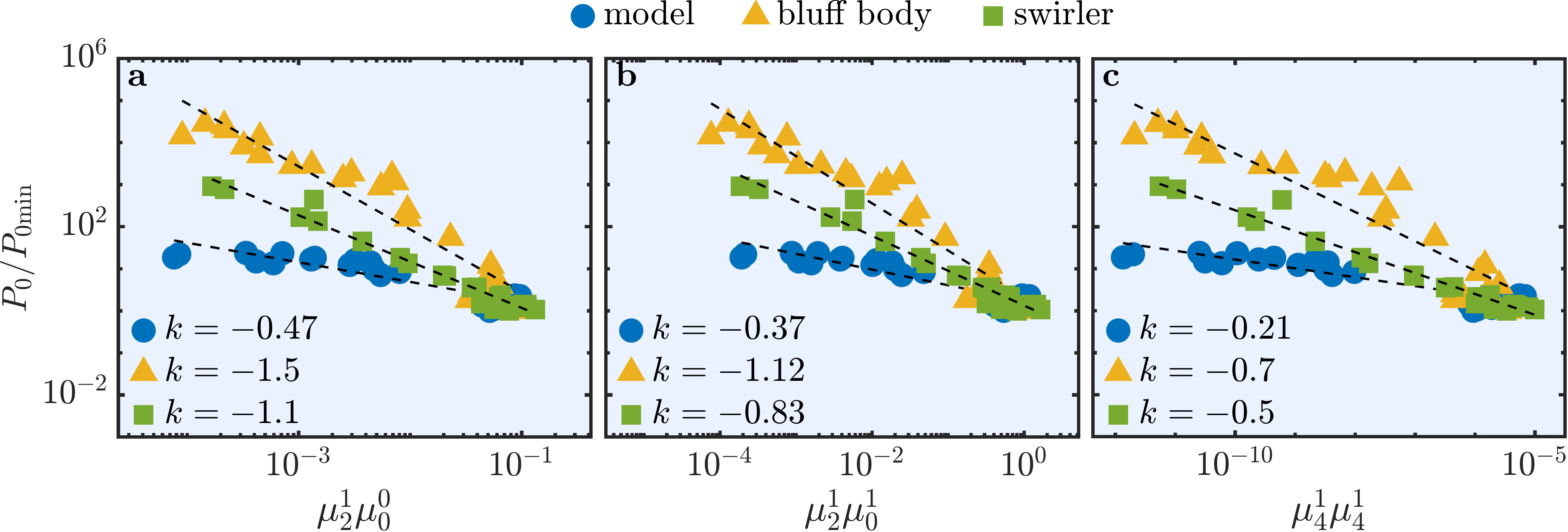}
\caption {Scaling behaviors of spectral condensation from the model and experiments, which follows $\log (P_0)~=~k \log ([\mu_{m}^{x} \mu_{n}^{y}]) +C$.
Note that both $P_0$ and $[\mu_{m}^{x} \mu_{n}^{y}]$ are plotted on a logarithmic scale, and $P_0$ is normalized by $P_{0 \rm min}$. Experimental data are from bluff-body-stabilized combustor by Pawar et al. \cite{pawar_2017_jfm}, and swirl-flow combustor by Nair et al.~\cite{nair2014jfm_intermittency}. From left to right: $\mu_{2}^{1} \mu_{0}^{0}$, $\mu_{2}^{1} \mu_{0}^{1}$, $\mu_{4}^{1} \mu_{4}^{1}$.}
\label{fig: 7}
\end{figure}

Next, we investigate the model's capability in capturing the self-organization using another measure reported in~\cite{pavithran2020epl, pavithran2021asme}. In the regime of combustion noise, the time series is aperiodic, with a low amplitude, and show a broadband profile in the power spectrum. However, when the system undergoes limit cycle oscillation due to thermoacoustic instability, the time series become periodic, the amplitude increases significantly, and the natural frequency shifts slightly and becomes dominant in the power spectrum. 
During the transition, the profile of the power spectrum narrows from a broad peak centered around the natural frequency to a single sharp peak (Fig.~\ref{fig: water_fall}). %\hl{Refer to waterfall figure}.
The measurement of spectral condensation quantifies how the power spectrum is condensing during the transition by calculating the $m^{th}$ moment of the power spectrum, evaluated as%as shown in Eq.~\ref{eq: spectral condensation}.

\begin{equation}
\left[\mu_{m}^{x} \mu_{n}^{y}\right]=\left[\int_{-\delta F}^{+\delta F} \frac{P(F)}{P_{0}}\left|\frac{F}{f_{0}}\right|^{m} d F\right]^{x} \times\left[\int_{-\delta F}^{+\delta F} \frac{P(F)}{P_{0}}\left|\frac{F}{f_{0}}\right|^{n} d F\right]^{y}
\label{eq: spectral condensation}
\end{equation}
$\mu_m$ here is the $m^{th}$ moment of the power spectrum, $f$ is the frequency, $f_0$ is the dominant frequency of time series, $P_0=P(f_0)$ is the dominant peak of the power spectrum, and $F=f-f_0$ is the modified frequency. $\left[\mu_{m}^{x} \mu_{n}^{y}\right]$ here is the products of higher-order moments of the power spectrum. In Fig.~\ref{fig: 7}, we are comparing a few of these high-order moment statistics obtained from the model and the experiments. The figure shows that for all the moments, the model and the experiments do follow a similar behavior in that during the transition from combustion noise to periodic oscillation, the variation of $\mu_{m}^{x} \mu_{n}^{y}$ follows a power law. 
The exponents of the power law are not the same constant across the model and the two experiments, primarily due to underlying coupling mechanisms. Nevertheless, the comparison confirms that the coupled oscillators can capture the underlying higher order statistics of spectral condensation observed during transition to the periodic states. 

\subsection{\textbf{Summary and outlook}}
We introduced a phenomenological model for modeling turbulent thermoacoustic systems. The model is based on the synchronization framework and considers the acoustic field and the unsteady heat release rate as two nonlinearly coupled sub-systems. 
The system consists of three oscillators, two of which form a sub-system representing the unsteady heat release rate from the turbulent reactive flow, and the third oscillator representing the acoustic field. 
The acoustic field and the unsteady release are coupled with a variable coupling strength.
By adjusting the coupling strength, the model can replicate the transition from stable operation to thermoacoustic instability in a turbulent combustor. In the stable operation regime, the model produces time series of fluctuations that has the characteristic of combustion noise. 
As the coupling strength increases, the system loses multifractality and transitions to periodic oscillation through intermittency. These behaviors are consistent with the experimental observations. 
Furthermore, we analyzed the self-organization behavior during the transition by calculating the scale of spectral condensation. 
The scale shows power laws similar to the ones reported in the experiments. 
In contrast to FTF/FDF, the developed model focuses on capturing the transition to thermoacoustic instability through intermittency. 
Currently, we have not tried to perform system identification and predict the unsteady regime or amplitude of oscillation in practical combustors.
However, the model provides a new configuration for modeling thermoacoustic systems, which considers the unsteady heat release rate as an independent sub-system, instead of a quasi-steady response to the acoustic perturbation. With the help of data-driven methods, the model can be further improved and optimized for real combustors.

\subsection*{\textbf{Acknowledgement}}
This work at UCSD is supported by the US National Science Foundation (Grant number: CBET-2053671). 
The authors are grateful to Dr. Samadhan A. Pawar and Dr. Vineeth Nair for sharing their experimental data for model validation. 
We thank Miss Induja Pavithran from the IIT Madras for useful discussions.
R. I. Sujith is grateful to the financial support of the Science and Engineering Research Board (SERB) of the Department of Science and Technology (DST) through its J. C. Bose Fellowship (JCB/2018/000034/SSC). 

% \subsection*{\textbf{Compliance with ethical standards}}

% \subsection*{\textbf{Conflict of interest}}
% The authors declare that they have no conflict of interest.

% \subsection*{\textbf{Data Availability}}
% The datasets generated during the current study are available from the corresponding authors on reasonable request.

\subsection*{\textbf{Appendix: Multifractal Detrended Fluctuation Analysis}}
In Multifractal Detrended Fluctuation Analysis (MFDFA), the signal of $x(i)$ will first be accumulated to generate a new signal (Eq.~\ref{eq: accum}), called "the profile".
\begin{equation}
\label{eq: accum}
Y_i~=~\sum_{k=1}^{i} (x_k-\bar x)
\end{equation}
Next, the profile with length $N$ is separated into $N_s$ equal-sized, non-overlapping segments.
the segments are detrended by subtracting the local linear fit $\bar Y$.
Then, a structure function $F_{s}^{w}$ of order $w$ and span $s$ can be obtained:
\begin{equation}
\label{eq: structure function}
F_{s}^{w}=\left(\frac{1}{N_{s}} \sum_{i=1}^{N_{s}}\left(\frac{1}{s} \sum_{i=1}^{s}\left(Y_{i}-\bar{Y}\right)^{2}\right)^{\frac{w}{2}}\right)^{\frac{1}{w}}
\end{equation}
The generalized Hurst exponent of order $w$ ($H^w$) is then defined as the slope of the linear regime of $F_{s}^{w}$ verses $s$ in a logarithmic plot.
Note that the Hurst exponent $H^2$ corresponds to the generalized Hurst exponent of order 2 (i. e. $w~=~2$). In this study, the range of $w$ is chosen from -2 to 2. 
From the  generalized Hurst exponents, we apply Legendre transformation to generate the singularity exponents $\alpha$ and the singularity spectrum $f(\alpha)$ by using the following relations:
% $$
% \tau_{w} =w H^{w}-1
% $$
% $$
% \alpha =\frac{\partial \tau_{w}}{\partial w}
% $$
% $$
% f(\alpha) =w \alpha-\tau_{w}
% $$

\begin{equation}
\tau_{w} =w H^{w}-1; ~~
\alpha =\frac{\partial \tau_{w}}{\partial w};~~
f(\alpha) =w \alpha-\tau_{w}
\end{equation}

% %\newpage
% \bibliographystyle{spmpsci_unsrt.bst}
\bibliography{paper.bib}
\bibliographystyle{ieeetr}
% \bibliographystyle{chicago}
% \spacingset{1}
% \bibliography{IISE-Trans}
	
\end{document}